\documentclass[aps,prl,twocolumn,groupedaddress,showpacs]{revtex4}

\usepackage{CJK}
\usepackage{graphicx}
\usepackage{amsfonts}
\usepackage{amsmath}
\usepackage{amssymb}
\usepackage{bm}

\begin{document}
\begin{CJK}{GBK}{song}
\title{Topological phase transitions with and without energy gap closing}
\author{Yunyou Yang$^1$}
\author{Huichao Li$^1$}
\author{L. Sheng$^1$}
\email{shengli@nju.edu.cn}
\author{R. Shen$^1$}
\author{D. N. Sheng$^2$}
\author{D. Y. Xing$^1$}
\email{dyxing@nju.edu.cn}
\affiliation{$^1$National Laboratory of Solid State Microstructures and
Department of Physics, Nanjing University, Nanjing 210093, China\\
$^2$ Department of Physics and Astronomy, California State
University, Northridge, California 91330, USA}
\date{\today }

\begin{abstract}

Topological phase transitions in a three-dimensional (3D)
topological insulator (TI) with an exchange field of strength $g$ are
studied by calculating spin Chern numbers
$C^\pm(k_z)$ with momentum $k_z$ as a parameter.
When $\vert g\vert$ exceeds a critical value $g_c$,
a transition of the 3D TI into a Weyl semimetal occurs,
where two Weyl points appear as critical points separating $k_z$ regions
with different first Chern numbers.
For $\vert g\vert<g_c$, $C^\pm(k_z)$ undergo a transition from $\pm
1$ to $0$ with increasing $\vert k_z\vert$ to a critical
value $k_z^{\mbox{\tiny C}}$. Correspondingly, surface states exist
for $\vert k_z\vert < k_z^{\mbox{\tiny C}}$, and vanish for $\vert
k_z\vert \ge k_z^{\mbox{\tiny C}}$. The transition at $\vert
k_z\vert = k_z^{\mbox{\tiny C}}$ is acompanied by closing of spin
spectrum gap rather than energy gap.
\end{abstract}

\pacs{72.25.-b, 73.23.-f,73.50.-h,73.20.At} \maketitle The quantum
Hall (QH) effect~\cite{iqhe,fqhe} in a two-dimensional (2D) electron
gas under a strong magnetic field provided the first example of
topological state of matter in condensed matter physics, which
cannot be described by the Landau theory of symmetry breaking.
Thouless, Kohmoto, Nightingale, and Nijs (TKNN) revealed that the
essential character of a QH insulator, different from an ordinary
insulator, is a topological invariant of occupied electron
states~\cite{tknn} or many-body wavefunctions~\cite{fchern}. They
related the Hall conductivity of the system to the first Chern
number (or TKNN number), which is quantized when the Fermi level
lies in an energy gap between  Landau levels. In such systems,
topological phase transitions can happen only by closing the energy
gap. Gapless edge states must appear on the boundary between a QH
insulator and an ordinary insulator, which is ensured by the
topological invariant. Interestingly, Haldane~\cite{haldane} 
proposed a spinless electron model on a 2D honeycomb lattice 
with staggered magnetic fluxes to realize the topological 
QH effect without  Landau levels.

The quantum spin Hall (QSH) effect was first theoretically predicted
by Kane and Mele~\cite{qshe1} and by Bernevig and
Zhang~\cite{qshe2}, and then experimentally observed in HgTe quantum
wells.~\cite{HgTe0,HgTe} Unlike the QH systems, where time-reversal
(TR) symmetry must be broken,  the QSH systems preserve the TR symmetry.
The main ingredient is the existence of strong spin-orbit coupling,
which acts as spin-dependent magnetic fluxes coupled to the electron
momentum. The QSH state is characterized by a bulk band gap and
gapless helical edge states on the sample
boundary.~\cite{qshe1,qshe2,HgTe0,HgTe,LSEdge,helical} The existence
of the edge states is due to nontrivial topological properties of
bulk energy bands. However, the bulk band topology of the QSH
systems cannot be classified by the first Chern number, which always
vanishes. Instead, it is classified by new topological invariants,
namely, the $Z_2$ index~\cite{Z2} or the spin Chern
numbers.~\cite{spinch1,spinch2,spinch3} For TR-invariant systems,
both $Z_2$ and spin Chern numbers were found to give an equivalent
description.~\cite{spinch2,spinch3} The robustness of the $Z_2$
index relies on the presence of the TR symmetry. In contrast, the
spin Chern numbers remain to be integer-quantized,
independent of any symmetry, as long as both the band gap and spin
spectrum gap stay open.~\cite{spinch2} They are also different from
the first Chern number for the QH state, which is protected by the
bulk energy gap alone. The spin Chern numbers have been employed to
study the TR-symmetry-broken QSH effect.~\cite{spinch4}

The QSH system is an example of the 2D topological insulators (TIs).
Its generalization to higher dimension led to the birth of 3D
TIs.~\cite{3dti1,3dti2,3DTI_Model,3d1,3d2} A 3D TI has a bulk band
gap and surface states on the sample boundary. The metallic surface
states provide a unique platform for realizing some exotic physical
phenomena, such as Majorana fermions~\cite{Majorana} and topological
magnetoelectric effect.~\cite{Magneto1,Magneto2} The 3D TIs have
been experimentally observed in Bi$_{1-x}$Sb$_x$, Bi$_2$Te$_3$, and
Bi$_2$Se$_3$
materials,~\cite{Exp3D1,Exp3D2,Exp3D3,Exp3D4,Exp3D5,Exp3D6} which
greatly stimulates the research in this field. The 3D TIs with TR
symmetry are usually classified by four $Z_2$
indices,~\cite{3dti1,3dti2} and are divided into two general classes:
strong and weak TIs, depending on the sum of the four $Z_2$ indices.
In the presence of disorder, while the weak TIs are unstable,  the
strong TIs remain to be robust. The $Z_2$ indices are essentially
defined only on the TR-symmetric planes in the Brillouin zone, and
do not provide information about the distribution of surface states
in the full momentum space. When the TR symmetry is broken, the
$Z_2$ indices become invalid. Therefore, a more general
characterization scheme for the bulk band topology, which does not
rely on any symmetry and can provide more information about the
distribution of surface states, is highly desirable.


In this Letter, for a 3D TI with an exchange field of strength $g$, we
consider a momentum component, e.g., $k_z$, as a parameter, and
analytically calculate spin Chern numbers $C^\pm(k_z)$ for the
effective 2D system. The phase diagram for $C^\pm(k_z)$ obtained
can describe the systematic evolution of the bulk band topology of the
3D TI with varying parameters, and provide more information
about the surface states. For small $g$, $C^\pm(k_z)$ take values $\pm
1$ for $\vert k_z\vert<k_z^{\mbox{\tiny C}}$, and undergo a
transition to $0$ at $\vert k_z\vert=k_z^{\mbox{\tiny C}}$.
Correspondingly, on a sample surface parallel to the $z$-axis,
helical surface states exist in the region $\vert
k_z\vert<k_z^{\mbox{\tiny C}}$, and disappear for $\vert k_z\vert
\ge k_z^{\mbox{\tiny C}}$. At $\vert k_z\vert = k_z^{\mbox{\tiny
C}}$, the spin spectrum gap closes, but the energy gap remains open,
which is distinct from a usual topological phase transition, where
the energy gap always collapses. When $\vert g\vert$ is greater than
a critical value $g_c$, a topological transition of the 3D TI into
a Weyl semimetal occurs. Two Weyl points appear as critical points
separating a QH phase of the effective 2D system for $\vert k_z\vert
<k_z^{\mbox{\tiny W}}$ from an ordinary insulator phase for $\vert
k_z\vert>k_z^{\mbox{\tiny W}}$, indicating that their appearance is
topological rather than accidental. Chiral surface states existing
in the region $\vert k_z\vert <k_z^{\mbox{\tiny W}}$ give rise to the
Fermi arcs.

Let us start from the effective Hamiltonian proposed in
Ref.~\cite{3DTI_Model}: $H=A_2\tau_x({k_x}{\sigma _x} + {k_y}{\sigma
_y}) +M({\bf k})\tau_z + {A_1}k_z\tau_x\sigma_z + g\sigma _z$, which
was used to describe the strong TI of Bi$_{2}$Se$_{3}$. Here,
 $\sigma_{m}$ and $\tau_{m}$ ($m=x,y$ or $z$)
denote the Pauli matrices in spin and orbital spaces, and $M({\bf
k})=M_0-B_1k_z^2-B_2k_{\perp}^2$ with $k_{\perp}^2=k_x^2+k_y^2$ is
the mass term expanded to the second order. In the last term, we
include an exchange field of strength $g$, in order to study the
TR-symmetry-broken effect on topological properties of the TI. For
convenience, the momentum is set to be dimensionless, by properly
choosing the units of parameters in the model, namely, $M_0$, $A_1$,
$A_2$, $B_1$, and $B_2$.

Making a unitary transformation ${\cal H}=U^\dagger HU$ with
$U=\frac{1}{2}(1+\tau_x)+\frac{1}{2} (1-\tau_x)\sigma_z$, we obtain
\begin{equation}
{\cal H}={A_2}({k_x}{\sigma _x} + {k_y}{\sigma _y})+[M({\bf k}){\tau
_z} + {A_1}{k_z}{\tau _x}+g]{\sigma _z}\ .
\end{equation}
The eigenstates of Hamiltonian (1) can be easily solved by first
diagonalizing the operator in the square bracket. The four
eigenenergies are obtained as
\begin{equation}
E^{v(c)}_{\pm}({\bf k})=-(+)\sqrt{A_2^2k_{\perp}^2+[g\pm\lambda({\bf
k})]^2}\ ,
\end{equation}
where $\lambda({\bf k})=\sqrt{M^2({\bf k})+A^2_{1}k_z^2}$, and
subscripts $\pm$ indicate two valence (conduction)  bands with
superscript $v$ ($c$). The electron wavefunctions in the valence
bands are given by
\begin{equation}
\varphi_\pm({\bf k})=\phi_\pm({\bf k})\otimes \chi_\pm({\bf k})\ .
\label{Eq7}
\end{equation}
Here, $\phi_{+}({\bf k})=[\mbox{sgn}(A_1k_z)\cos{\alpha_k}$,
$\sin{\alpha_k}]^{T}$ and $\phi_{-}({\bf
k})=[\mbox{sgn}(A_1k_z)\sin{\alpha_k}$, $-\cos{\alpha_k}]^{T}$ are
wavefunctions in the $\tau_z$ space, and $\chi_{\pm}=(\frac{k_x -
ik_y}{k_{\perp}}\sin{\theta_{k}^{\pm}}$,
$-\cos{\theta_{k}^{\pm}})^{T}$ are wavefunctions in the $\sigma_z$
space, with $2\alpha_k=\mbox{ctg}^{-1}[{M({\bf k})}/{\vert
A_1k_z\vert}]$ and $2\theta^{\pm}_{k}=
\mbox{ctg}^{-1}[(g\pm\lambda({\bf k}))/{\vert A_2k_{\perp}\vert}]$.

The basic idea of our theoretical calculation is explained as
follows. We consider one of the momentum components, e.g., $k_z$ as
a parameter. For a given $k_z$, Eq.\ (1) is equivalent to a 2D
system, for which spin Chern numbers $C^{\pm}(k_z)$ can be defined.
For a semi-infinite sample of the 3D TI with its surface parallel to
the $z$ axis, $k_z$ remains to be a good quantum number.
Correspondingly, nonzero $C^{\pm}(k_z)$ indicate that edge states
with the given $k_z$ must appear on the edge of the effective 2D
system. The edge states at various $k_z$ essentially form  surface
states of the 3D sample. Therefore, the characteristics of the
surface states can be determined from the calculation of the
$k_z$-dependent spin Chern numbers $C^{\pm}(k_z)$.

The spin Chern numbers for the effective 2D system are calculated in
a standard way, which has been described in details in previous
works.~\cite{spinch3,spinch4} By studying a special case of $k_z=0$,
we find that the topological properties of Eq.\ (1) can be described
by the spin Chern numbers $C^{\pm}(k_z)$ associated with $\tau_z$.
Here, $\tau_z$ corresponds to $U\tau_zU^\dagger=\tau_z\sigma_z$ in
the original Hamiltonian $H$, and so can be considered as a spin
operator, measuring the difference of spin polarization between the
two orbits. The eigenstates of projected spin operator $P\tau_z P$
need to be calculated first, where $P$ is the projection operator
into the valence bands. Since $P\tau_z P$ commutes with momentum
operator, its eigenstates can be obtained at each momentum ${\bf k}$
separately. The eigenvalues of $P\tau_z P$ are given by
\begin{equation}
\xi^{\pm}({\bf k})=\pm\sqrt{\cos^2 2\alpha_k+\cos^2
(\theta_{k}^{+}-\theta_{k}^{-})\sin^2 2\alpha_{k}}\ .
\end{equation}
The corresponding eigenfunctions are denoted by $\Psi^{\pm}({\bf
k})$, whose expressions are lengthy and will not be written out
here. The spin Chern numbers are just the Chern numbers of the two
spin sectors formed by $\Psi^{\pm}({\bf k})$, i.e.,
$C^{\pm}(k_z)=\frac{i}{2\pi}\int dk_xdk_y\hat{\bf
e}_z\cdot\bigl[\nabla_2\times\langle \Psi^{\pm}({\bf
k})\vert\nabla_{2}\vert\Psi^{\pm}({\bf k}\rangle\bigr]$, where
$\nabla_2$ is the 2D Laplace operator acting on ($k_x$, $k_y$). By
some algebra, $C^{\pm}(k_z)$ are derived to be
\begin{eqnarray}
C^\pm(k_z)=\pm\frac{1}{2}\Big[\mbox{sgn}(B_2)+\mbox{sgn}\Big(Q(k_z)\mbox{\rm sgn} [P(k_z)]\pm g\Big)\Big]
\label{Eq16}
\end{eqnarray}
with $P(k_z)=M_0-B_1k_z^2$ and $Q(k_z)=\sqrt{P^2(k_z)+A_1^2k_z^2}$.

\begin{figure}
\includegraphics[width=2.0in]{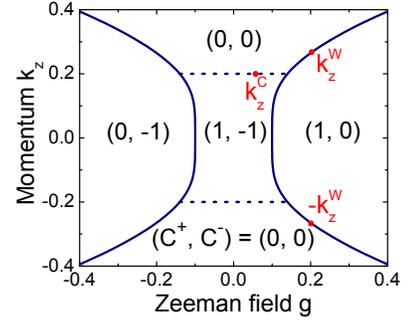}
\caption{Phase diagram determined from  spin Chern numbers in the
$k_z$-$g$ plane. The parameters are chosen to be $M_0=0.1$,
$A_1=0.7$, $A_2=1$, $B_1=2.5$, and $B_2=14$.} \label{Fig1}
\end{figure}
Equations\ (2), (4), and (5) are the main analytical results of this
work. For $g=0$, the spin Chern numbers at $k_z=0$ reduces to
$C^{\pm}(0)=\pm\frac{1}{2}[\mbox{sgn}(B_2)+\mbox{sgn}(M_0)]$.
$C^{\pm}(0)$ are nonzero when $B_2$ and $M_0$ have the same sign,
and vanish otherwise. $C^{\pm}(0)$ play a role similar to the $Z_2$
index. Nonzero $C^{\pm}(0)$ ensure that surface states exist in the
vicinity of $k_z=0$ on a surface parallel to
the $z$ axis.  Without loss of generality,  
we will focus on the parameter region of $B_2>0$ and $M_0>0$, to
which Bi$_2$Se$_3$ belongs. We wish to emphasize here that when
$k_z$ is considered as a parameter, the effective 2D Hamiltonian (1)
breaks the TR symmetry for any $k_z\neq 0$, even if $g=0$, as its TR
couterpart is at $-k_z$. Therefore, while the $k_z$-dependent spin
Chern numbers given by Eq.\ (5) remain to be valid at any $k_z$, a
$Z_2$ index cannot be defined for any $k_z\neq 0$. Eq.\ (5) allows
us to extract more information about the basic characteristics of
the surface states.

A typical phase diagram for the spin Chern numbers in the $k_z$
versus $g$ plane, as determined by Eq.\ (5), is plotted in Fig.\ 1.
For simplicity, $A_2$ is taken to be the unit of energy. For small
$\vert g\vert$, $C^{\pm}(k_z)=\pm 1$ at small $\vert k_z\vert$,
corresponding to a QSH phase of the effective 2D system, and drop to
$0$ with increasing $\vert k_z\vert$ to a critical value
$k_{z}^{\mbox{\tiny C}}=\sqrt{M_0/B_1}$, as indicated by the dotted
lines. The system becomes an ordinary insulator for $\vert k_z\vert
> k_{z}^{\mbox{\tiny C}}$. When $\vert g\vert$ is greater than a
critical value $g_c$, the effective 2D system enters a QH phase with
a nonzero total (first) Chern number $C(k_z)\equiv C^{+}(k_z)+C^{-}(k_z)=1$
if $g>0$, and $-1$ if $g<0$. The boundary
enclosing this phase is determined by equation $g=\pm Q(k_z)$, as
indicated by the solid curves. The critical exchange field is given by
$g_c=\mbox{min}[Q(k_z)]$, and $g_c\simeq 0.1$ for the parameters used 
in Fig.\ 1. 

\begin{figure}
\includegraphics[width=2.0in]{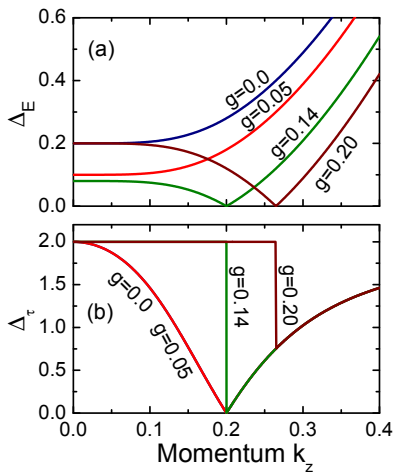}
\caption{Energy gap $\Delta_{E}$ (a) and  spin spectrum gap
$\Delta_\tau$ (b)
         as functions of momentum $k_z$ for different values of $g$. The other parameters
         are the same as in Fig.\ 1.
       }
\label{Fig2}
  \end{figure}
It is interesting to see how the energy gap
$\Delta_{E}(k_z)=\mbox{min}[E^{c}_{\pm}({\bf k}) -E^{v}_{\pm}({\bf
k})]\vert_{k_z}$ and the spin spectrum gap
$\Delta_\tau(k_z)=\mbox{min}[\xi^{+}({\bf k}) -\xi^{-}({\bf
k})]\vert_{k_z}$ behave on the boundary (solid and dashed lines)
between different phases. From Eqs.\ (2) and (4), we find that on
the dotted lines in Fig.\ 1, the spin spectrum gap closes at
$k_x=k_y=0$, but the energy gap remains open. On the contrary, on
the solid boundary lines, the energy gap closes at $k_x=k_y=0$, but
the spin spectrum gap remains open. $\Delta_E$ and $\Delta_\tau$ as
functions of $k_z$ for several values of $g$ are plotted in Figs.\
2(a) and 2(b), respectively.  We notice that at  $g=0.14$ and
$k_z=k_z^{\mbox{\tiny C}}=0.2$, the energy gap and spin spectrum gap
vanish simultaneously. This is because the dotted and solid boundary
lines in Fig.\ 1 intersect just at that point.

\begin{figure}
\includegraphics[width=3.2in]{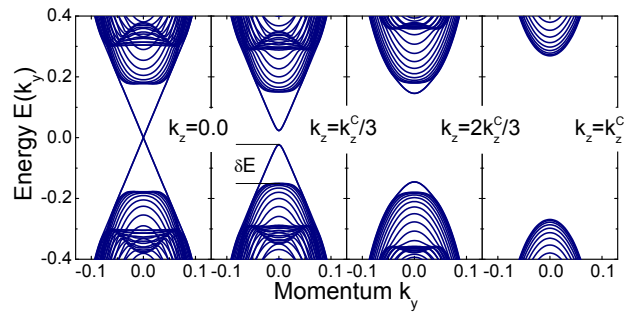}
\caption{Profiles of energy spectrum for a semi-finite
    sample of the 3D TI for four different $k_z$ at $g=0.05$.
    The other parameters are the same as in Fig.\ 1.
    }
\label{Fig3}
\end{figure}
\begin{figure}
\includegraphics[width=3.0in]{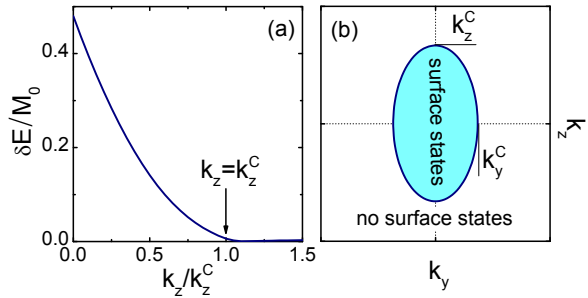}
\caption{(a) Maximum level spacing $\delta E$ between
the edge states and bulk states
as a function of $k_z/k_z^{\mbox{\tiny C}}$. (b) A schematic of the distribution of
surface states in 2D momentum space.}
\label{Fig4}
\end{figure}
To study the surface states directly, we construct a tight binding
model on a cubic lattice with two spins and two orbits on each site,
which recovers the Hamiltonian Eq.\ (1) in the continuum limit. A
semi-infinite sample with its surface parallel to the $y$-$z$ plane
is considered, where $k_y$ and $k_z$ remain to be good quantum
numbers. The calculated energy spectrum for $g=0.05$ is plotted as a
function of $k_y$ for four different values of $k_z$ in Fig.\ 3.
Although $g\neq 0$ breaks the TR symmetry, the surface states remain
to be gapless at $k_z=0$, because $\tau_z$ in Eq.\ (1) is conserved
at $k_z=0$. From Fig.\ 3, it is found that for $k_z=0$,
$k_{z}^{\mbox{\tiny C}}/3$, and $2k_{z}^{\mbox{\tiny C}}/3$, surface
states always exist in the bulk energy gap, but no surface
states appear at $k_z=k_z^{\mbox{\tiny C}}$. To see the evolution of surface
states with $k_z$ more clearly, we define a maximum level spacing
$\delta E$ between the surface states and bulk states, as
illustrated in Fig.\ 3. In Fig.\ 4(a), $\delta E$ is plotted as a
function of $k_z$. One can see that $\delta E$ decreases with
increasing $k_z$, and drops to nearly $0$ at $k_z\ge
k_z^{\mbox{\tiny C}}$. Therefore, we conclude that surface states
exist only in the region $\vert k_z\vert<k_z^{\mbox{\tiny C}}$, and
vanish for $\vert k_z\vert\ge k_z^{\mbox{\tiny C}}$, which is well
consistent with the phase diagram of Fig.\ 1.

The critical momentum $k_z^{\mbox{\tiny C}}$ marks a topological
phase transition, characterized by the change in the spin Chern
numbers, with disappearance of surface states as an observable
consequence. This topological phase transition is unconventional,
which is accompanied by closing the spin spectrum gap rather than
the energy gap. While the closing of the spin spectrum gap may not
be observed directly, we find that the average of operator $\tau_z$
in $H$, namely, $\sum_{\beta=\pm}\langle\varphi_{\beta}({\bf
k})\vert\ U^\dagger\tau_zU\vert\varphi_{\beta}({\bf
k})\rangle\propto\cos(2\alpha_{k})$, changes its sign at ${\bf
k}=(0, 0, \pm k_z^{\mbox{\tiny C}})$, which indicates a reversal of
the orbital polarization, leading to a bulk physical observable at
the transition. For the strong TI under consideration, the
nontrivial bulk band topology can be examined in any direction. For
example, by considering $k_y$ as a parameter, we can calculate the
spin Chern numbers in the effective 2D space of $(k_x, k_z)$, and
obtain a critical $k_y^{\mbox{\tiny C}}=\sqrt{M_0/B_2}$. Combining
$k_y^{\mbox{\tiny C}}$ and $k_z^{\mbox{\tiny C}}$ together, we can
depict the region in the $k_y$-$k_z$ plane, where topological
surface states exist on a surface parallel to the $y$-$z$ plane, as
shown in Fig.\ 4(b).

\begin{figure}
\includegraphics[width=3.2in]{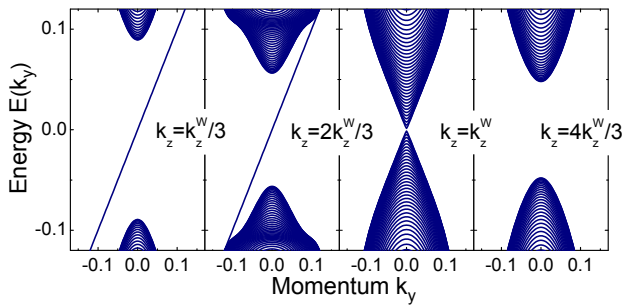}
\caption{Profiles of energy spectrum of a semi-finite
    sample for four different $k_z$ at $g=0.2$.
    The other parameters are the same as in Fig.\ 1.
    }
\label{Fig5}
\end{figure}
For $\vert g\vert>g_c$, the 3D system enters another topological
phase characterized by a nonzero total Chern number $C(k_z)$ for
small $\vert k_z\vert$, which is essentially a Weyl semimetal
phase.~\cite{FermiArc1,FermiArc2,FermiArc3,FermiArc4} The quantum
phase transition from the 3D TI to the Weyl semimetal with tunning
$g$ can be understood as a topological transition, at which the
energy gap closing causes one of spin Chern numbers $C^{\pm}(k_z)$
to vanish, while the other remains to be quantized. At a given $g$,
from equation $g=\pm Q(k_z)$ for the solid boundary lines in Fig.\
1, one can obtain two critical values of $k_z$, namely,
$\pm k_z^{\mbox{\tiny W}}$, as indicated in Fig.\ 1. ${\bf k}=(0, 0,\pm
k_z^{\mbox{\tiny W}})$ are a pair of Weyl points, or 3D Dirac
points, at which the conduction and valence bands touch. The
appearance of Weyl points is usually attributed to accidental
degeneracy or
symmetry.~\cite{FermiArc1,FermiArc2,FermiArc3,FermiArc4} The
obtained phase diagram in Fig.\ 1 sheds more light on its
topological origin. At a given $g$, the Weyl points appear as two
critical points separating a QH state of the effective 2D system for
$\vert k_z\vert<k_z^{\mbox{\tiny W}}$ from an ordinary insulator
state for $\vert k_z\vert>k_z^{\mbox{\tiny W}}$, at which the energy
gap must vanish. Changing parameters cannot open energy gaps at the
two Weyl points, unless they come together, so that the QH state in
between is destroyed. At $g=0.2$, the calculated energy spectra for
four different $k_z$ as functions of $k_y$  are plotted in Fig.\ 5
for a semi-infinite sample with its surface parallel to the $y$-$z$
plane. It is found that for $k_z<k_z^{\mbox{\tiny W}}$, chiral
surface states appear in the energy gap; the conduction and valence
bands touch at $k_z=k_z^{\mbox{\tiny W}}$; and  for
$k_z>k_z^{\mbox{\tiny W}}$, the energy gap reopens, but surface
states no longer exist. These results are in good agreement with the
phase diagram in Fig.\ 1. The chiral surface states appearing in the
region of $\vert k_z\vert<k_z^{\mbox{\tiny W}}$ give rise to the
Fermi arcs.~\cite{FermiArc1,FermiArc2,FermiArc3,FermiArc4} The Weyl semimetals
with Fermi arcs have been paid much attention recently. 
Our work suggests that a possible route to realize such an interesting
3D topological state of matter is through magnetic doping in 3D TIs.



This work is supported by the State Key Program for Basic Researches
of China under Grants Nos. 2009CB929504 (LS),
2011CB922103, and 2010CB923400 (DYX), the National Natural Science
Foundation of China under Grant Nos. 11225420, 11074110 (LS), 11074111 (RS),
11174125, 11074109, 91021003 (DYX),
and a project funded by the PAPD
of Jiangsu Higher Education Institutions. We also thank the US NSF Grants
No. DMR-0906816 and No. DMR-1205734, and
Princeton MRSEC Grant No. DMR-0819860 (DNS).


\end{CJK}
\end{document}